\documentclass{revtex4}
\usepackage{amssymb}
\usepackage{amsfonts}
\usepackage{amsmath}

\input{tcilatex}

\begin{document}

\title{Comment on "Static and spherically symmetric black holes in f(R)
theories"}
\author{S. Habib Mazharimousavi}
\email{habib.mazhari@emu.edu.tr}
\author{M. Halilsoy}
\email{mustafa.halilsoy@emu.edu.tr }
\affiliation{Department of Physics, Eastern Mediterranean University, G. Magusa, north
Cyprus, Mersin 10, Turkey. }

\begin{abstract}
We consider the interesting "near-horizon test" reported in (PRD84,
084006(2011)) for any static, spherically symmetric (SSS) black hole
solution admitted in f(R) gravity. Before adopting the necessary conditions
for the test, however, revisions are needed as we point out in this Comment.
\end{abstract}

\maketitle

In order to derive the necessary conditions for the existance of static,
spherically symmetric (SSS) black holes we consider the series expansions of
all expressions in the vicinity of the event horizon \cite{1}.

Our $4-$dimensional action which represents the Einstein- $f\left( R\right) $
gravity is given by%
\begin{equation}
S=\frac{1}{2\kappa }\int \sqrt{-g}f\left( R\right) d^{4}x
\end{equation}%
in which $\kappa =8\pi G,$ and $f\left( R\right) $ is a real arbitrary
function of the Ricci scalar $R.$ The $4-$dimensional SSS black hole's line
element is chosen to be as \cite{1}%
\begin{equation}
ds^{2}=-e^{-2\Phi }\left( 1-\frac{b}{r}\right) dt^{2}+\frac{1}{\left( 1-%
\frac{b}{r}\right) }dr^{2}+r^{2}\left( d\theta ^{2}+\sin ^{2}\theta d\varphi
^{2}\right) 
\end{equation}%
where $\Phi $ and $b$ are two unknown real functions of $r$ and at the
horizon, $r=r_{0},$ we have $b\left( r_{0}\right) =b_{0}=r_{0}.$

Variation of the action with respect to the metric gives the following field
equations%
\begin{equation}
f_{R}R_{\mu }^{\nu }-\frac{f}{2}\delta _{\mu }^{\nu }-\nabla ^{\nu }\nabla
_{\mu }f_{R}+\delta _{\mu }^{\nu }\square f_{R}=0
\end{equation}%
in which $\square =\nabla ^{\mu }\nabla _{\mu }=\frac{1}{\sqrt{-g}}\partial
_{\mu }\left( \sqrt{-g}\partial ^{\mu }\right) $ and $\nabla ^{\nu }\nabla
_{\mu }h=g^{\lambda \nu }\nabla _{\lambda }h_{,\mu }=g^{\lambda \nu }\left(
\partial _{\lambda }h_{,\mu }-\Gamma _{\lambda \mu }^{\beta }h_{,\beta
}\right) $ \cite{2,3}. This leads to the field equations%
\begin{equation}
f_{R}R_{t}^{t}-\frac{f}{2}+\square f_{R}=\nabla ^{t}\nabla _{t}f_{R}
\end{equation}%
\begin{equation}
f_{R}R_{r}^{r}-\frac{f}{2}+\square f_{R}=\nabla ^{r}\nabla _{r}f_{R}
\end{equation}%
\begin{equation}
f_{R}R_{\theta }^{\theta }-\frac{f}{2}+\square f_{R}=\nabla ^{\theta }\nabla
_{\theta }f_{R}
\end{equation}%
which are independent. Note that the $\varphi \varphi $ equation is
identical with $\theta \theta $ equation. By adding the four equations
(i.e., $tt$, $rr$, $\theta \theta $ and $\varphi \varphi $) we find 
\begin{equation}
f_{R}R-2f+3\square f_{R}=0
\end{equation}%
which is the trace of Eq. (3). We note that this is not an independent
equation from the other three equations. One may consider this equation with
only two of the others. In other words, if one considers the latter equation
with the other three equations two of them become identical. The Eq.s (3-6)
of Ref. \cite{1} involve unfortunate errors so that we evaluate each Ricci
tensor component in some detail. In the following we shall expand the
unknown functions about the horizon which will determine the near horizon
behaviour. To do so we introduce 
\begin{equation}
\epsilon x=r-r_{0},\text{ \ \ \ \ \ }\left\vert \epsilon \right\vert \ll 1
\end{equation}%
so that 
\begin{eqnarray}
\Phi  &=&\Phi _{0}+\Phi _{0}^{\prime }\epsilon x+\frac{1}{2}\Phi
_{0}^{\prime \prime }\epsilon ^{2}x^{2}+... \\
b &=&b_{0}+b_{0}^{\prime }\epsilon x+\frac{1}{2}b_{0}^{\prime \prime
}\epsilon ^{2}x^{2}+... \\
f &=&f_{0}+f_{0}^{\prime }\epsilon x+\frac{1}{2}f_{0}^{\prime \prime
}\epsilon ^{2}x^{2}+... \\
R &=&R_{0}+R_{0}^{\prime }\epsilon x+\frac{1}{2}R_{0}^{\prime \prime
}\epsilon ^{2}x^{2}+.... \\
F &=&\frac{df}{dR}=F_{0}+F_{0}^{\prime }\epsilon x+\frac{1}{2}F_{0}^{\prime
\prime }\epsilon ^{2}x^{2}+... \\
E &=&\frac{d^{2}f}{dR^{2}}=E_{0}+E_{0}^{\prime }\epsilon x+\frac{1}{2}%
E_{0}^{\prime \prime }\epsilon ^{2}x^{2}+... \\
H &=&\frac{d^{3}f}{dR^{3}}=H_{0}+H_{0}^{\prime }\epsilon x+\frac{1}{2}%
H_{0}^{\prime \prime }\epsilon ^{2}x^{2}+...
\end{eqnarray}%
in which a prime denotes derivative with respect to $r.$ Other notation is
such that $Y_{0}=Y\left( r_{0}\right) ,$ $Y_{0}^{\prime }=\left. \frac{dY}{dr%
}\right\vert _{r=r_{0}},$ $Y_{0}^{\prime \prime }=\left. \frac{d^{2}Y}{dr^{2}%
}\right\vert _{r=r_{0}}$and so on, in which $Y$ represents any function used
here. We evaluate also the near horizon form of $\square f_{R}$ and the
other similar terms:%
\begin{equation}
\square f_{R}=\frac{d^{3}f}{dR^{3}}g^{11}\left( R^{\prime }\right) ^{2}+%
\frac{d^{2}f}{dR^{2}}\square R
\end{equation}%
where 
\begin{equation}
\square R=\left( R^{\prime \prime }+\frac{R^{\prime }}{r}-\Phi ^{\prime
}R^{\prime }\right) \left( 1-\frac{b}{r}\right) +\frac{R^{\prime }}{r}\left(
1-b^{\prime }\right) 
\end{equation}%
which, up to the first order in $\epsilon $ it would read%
\begin{equation*}
\square f_{R}\simeq -\frac{E_{0}R_{0}^{\prime }\left( b_{0}^{\prime
}-1\right) }{r_{0}}-\frac{\left( H_{0}R_{0}^{\prime 2}-\left( E_{0}\Phi
_{0}^{\prime }-E_{0}^{\prime }\right) R_{0}^{\prime }+2E_{0}R_{0}^{\prime
\prime }\right) \left( b_{0}^{\prime }-1\right) +R_{0}^{\prime
}E_{0}b_{0}^{\prime \prime }}{r_{0}}\epsilon x.
\end{equation*}%
Similarly, the other terms to the first order read:%
\begin{eqnarray}
\nabla ^{t}\nabla _{t}f_{R} &=&-g^{tt}\frac{d^{2}f}{dR^{2}}R^{\prime }\Gamma
_{tt}^{r}\simeq  \\
&&-\frac{E_{0}R_{0}^{\prime }\left( b_{0}^{\prime }-1\right) }{2r_{0}}-\frac{%
r_{0}E_{0}R_{0}^{\prime }b_{0}^{\prime \prime }+\left( b_{0}^{\prime
}-1\right) \left[ E_{0}^{\prime }R_{0}^{\prime }r_{0}+E_{0}\left(
R_{0}^{\prime \prime }r_{0}-2R_{0}^{\prime }\left[ 1+\Phi _{0}^{\prime }r_{0}%
\right] \right) \right] }{2r_{0}^{2}}\epsilon x,  \notag
\end{eqnarray}%
\begin{eqnarray}
\nabla ^{r}\nabla _{r}f_{R} &=&g^{rr}\left( R^{\prime \prime }\frac{d^{2}f}{%
dR^{2}}+R^{\prime 2}\frac{d^{3}f}{dR^{3}}-\frac{d^{2}f}{dR^{2}}R^{\prime
}\Gamma _{tt}^{r}\right) \simeq  \\
&&-\frac{E_{0}R_{0}^{\prime }\left( b_{0}^{\prime }-1\right) }{2r_{0}}-\frac{%
r_{0}E_{0}R_{0}^{\prime }b_{0}^{\prime \prime }+\left( b_{0}^{\prime
}-1\right) \left[ R_{0}^{\prime }r_{0}\left( 2R_{0}^{\prime
}H_{0}+E_{0}^{\prime }\right) -E_{0}\left( 2R_{0}^{\prime
}-3r_{0}R_{0}^{\prime \prime }\right) \right] }{2r_{0}^{2}}\epsilon x  \notag
\end{eqnarray}%
and 
\begin{equation}
\nabla ^{\theta }\nabla _{\theta }f_{R}=-g^{\theta \theta }\frac{d^{2}f}{%
dR^{2}}R^{\prime }\Gamma _{\theta \theta }^{r}\simeq -\frac{%
E_{0}R_{0}^{\prime }\left( b_{0}^{\prime }-1\right) }{r_{0}^{2}}\epsilon x.
\end{equation}%
Accordingly, the Ricci tensor components become:%
\begin{equation}
R_{t}^{t}=\frac{b_{0}^{\prime \prime }-3\left( b_{0}^{\prime }-1\right) \Phi
_{0}^{\prime }}{2r_{0}}+\frac{\left( b_{0}^{\prime }-1\right) \left[ \left(
2\Phi _{0}^{\prime 2}-5\Phi _{0}^{\prime \prime }\right) r_{0}+2\Phi
_{0}^{\prime }\right] -b_{0}^{\prime \prime }\left( 3\Phi _{0}^{\prime
}r_{0}+1\right) +r_{0}b_{0}^{\prime \prime \prime }}{2r_{0}^{2}}\epsilon x
\end{equation}%
\begin{equation}
R_{r}^{r}=R_{t}^{t}+\frac{2\left( b_{0}^{\prime }-1\right) \Phi _{0}^{\prime
}}{r_{0}^{2}}\epsilon x
\end{equation}%
and finally%
\begin{equation}
R_{\varphi }^{\varphi }=R_{\theta }^{\theta }=\frac{b_{0}^{\prime }}{r_{0}}-%
\frac{\left( b_{0}^{\prime }-1\right) \left( \Phi _{0}^{\prime
}r_{0}+2\right) -b_{0}^{\prime \prime }r_{0}+2}{r_{0}^{3}}\epsilon x.
\end{equation}%
Next, we rewrite the field equations up to the first order in $\epsilon $.
After matching the zeroth order terms we find from (4) and (5) 
\begin{equation}
-\left( b_{0}^{\prime }-1\right) \left( 3\Phi _{0}^{\prime
}F_{0}+E_{0}R_{0}^{\prime }\right) +F_{0}b_{0}^{\prime \prime }-f_{0}r_{0}=0,
\end{equation}%
while from (6) it yields%
\begin{equation}
\left( b_{0}^{\prime }-1\right) \left( F_{0}-E_{0}R_{0}^{\prime
}r_{0}\right) +F_{0}-\frac{1}{2}f_{0}r_{0}^{2}=0.
\end{equation}%
From these equations one finds%
\begin{equation}
\frac{f_{0}}{F_{0}}=-2\frac{\left( 3\Phi _{0}^{\prime }r_{0}+1\right) \left(
b_{0}^{\prime }-1\right) +1-b_{0}^{\prime \prime }r_{0}}{r_{0}^{2}}
\end{equation}%
and 
\begin{equation}
\frac{F_{0}}{E_{0}}=\frac{r_{0}\left( b_{0}^{\prime }-1\right) R_{0}^{\prime
}}{6b_{0}^{\prime }-3\Phi _{0}^{\prime }r_{0}\left( b_{0}^{\prime }-1\right)
+b_{0}^{\prime \prime }r_{0}-2R_{0}r_{0}^{2}}.
\end{equation}%
Knowing that the explicit form of $R_{0}$ and $R_{0}^{\prime }$ are given by%
\begin{equation}
R_{0}=\frac{2+b_{0}^{\prime \prime }r_{0}+\left( 2-3\Phi _{0}^{\prime
}r_{0}\right) \left( b_{0}^{\prime }-1\right) }{r_{0}^{2}},
\end{equation}%
and 
\begin{equation}
R_{0}^{\prime }=\frac{\left[ \left( 2\Phi _{0}^{\prime 2}-5\Phi _{0}^{\prime
\prime }\right) r_{0}^{2}+2\Phi _{0}^{\prime }r_{0}-4\right] \left(
b_{0}^{\prime }-1\right) -r_{0}^{2}\left( 3\Phi _{0}^{\prime }b_{0}^{\prime
\prime }-b_{0}^{\prime \prime \prime }\right) -4+b_{0}^{\prime \prime }r_{0}%
}{r_{0}^{3}}.
\end{equation}%
Unlike the result of Eq. (14) Ref. \cite{1}, here from Eq.s (26) and (27) we
obtain%
\begin{equation}
\frac{f_{0}}{F_{0}}=2R_{0}-\frac{6b_{0}^{\prime }}{r_{0}^{2}},\text{ \ and \ 
}\frac{F_{0}}{E_{0}}=\frac{R_{0}^{\prime }r_{0}\left( b_{0}^{\prime
}-1\right) }{4b_{0}^{\prime }-R_{0}r_{0}^{2}}.
\end{equation}%
On the other hand, from (11)-(15), one finds 
\begin{equation}
F_{0}=\frac{f_{0}^{\prime }}{R_{0}^{\prime }};\text{ \ }E_{0}=\frac{%
F_{0}^{\prime }}{R_{0}^{\prime }};\text{ \ \ }H_{0}=\frac{E_{0}^{\prime }}{%
R_{0}^{\prime }}.
\end{equation}%
As we mentioned before, Eq. (7) is not independent and is identically
satisfied. After the zeroth order terms one may look at the first order
equations which upon combination of Eq.s (4) and (5) we get%
\begin{equation}
r_{0}\left( H_{0}R_{0}^{\prime 2}+E_{0}R_{0}^{\prime \prime
}+E_{0}R_{0}^{\prime }\Phi _{0}^{\prime }\right) +2\Phi _{0}^{\prime }F_{0}=0
\end{equation}%
which consequently implies%
\begin{equation}
\frac{H_{0}}{E_{0}}=\frac{\left( 4R_{0}^{\prime \prime }+6R_{0}^{\prime
}\Phi _{0}^{\prime }\right) \left( b_{0}^{\prime }-1\right) +\left(
R_{0}^{\prime \prime }+R_{0}^{\prime }\Phi _{0}^{\prime }\right) \left(
4-R_{0}r_{0}^{2}\right) }{\left( R_{0}r_{0}^{2}-4b_{0}^{\prime }\right)
R_{0}^{\prime 2}}.
\end{equation}%
The other equations are rather complicated so that we will not write them
openly.

Our conclusion for a general SSS black hole solution in $f(R)$ gravity is
that the necessary conditions for an $f(R)$ gravity to have $b_{0}=r_{0}$
type of solution are given by (28)-(31). To have a Schwarzschild-like black
hole with $b(r)=M=$constant, these conditions reduce to the simpler
condition as%
\begin{equation}
\frac{f_{0}}{F_{0}}=2R_{0}
\end{equation}%
which for example in $f(R)=\alpha \left( R+\beta \right) ^{n},$ with $\alpha
,$ $\beta $ constants, metric is viable only for $n=\frac{1}{2}.$ We add
that for the Schwarzschild case (i.e. $n=1,$ $\alpha =1,$ $\beta =0$)
condition (34) is trivially satisfied so that the "near-horizon test"
concerns non-Schwarzschild SSS metrics in $f(R)$ gravity. Our result is also
in conform with Ref. \cite{4}.

\bigskip

\bigskip

\bigskip

\end{document}